\documentclass[a4paper]{JHEP}

\usepackage{graphicx}
\usepackage{amssymb}
\usepackage{latexsym}
\usepackage[mathscr]{eucal}
\usepackage{amsmath}
\usepackage{cite}
\usepackage{afterpage}

\newcommand{\bgam}{\beta_\gamma}
\newcommand{\bmat}{\beta_\mathrm{m}}
\newcommand{\delphi}{\delta\phi}
\newcommand{\Mpl}{M_\mathrm{Pl}}

\newcommand{\gsimm}{\raise.3ex\hbox{$>$\kern-.75em\lower1ex\hbox{$\sim$}}}
\newcommand{\lsimm}{\raise.3ex\hbox{$<$\kern-.75em\lower1ex\hbox{$\sim$}}}

\newcommand{\be}{\begin{equation}}
\newcommand{\ee}{\end{equation}}
\newcommand{\ba}{\begin{eqnarray}}
\newcommand{\ea}{\end{eqnarray}}

\newcommand{\bea}{\begin{eqnarray*}}
\newcommand{\eea}{\end{eqnarray*}}

\title{Chameleon Fragmentation }

\author{Philippe Brax \\
  Institut de Physique Th\'eorique, CEA, IPhT, CNRS, URA 2306,
  F-91191Gif/Yvette Cedex, France \\ E-mail:
  \email{philippe.brax@cea.fr}}

\author{Amol Upadhye \\
  Institute for the Early Universe, Ewha University,
  Seoul 120-750 Korea;\\
  Lawrence Berkeley National Laboratory,
  University of California, Berkeley, CA 94720, USA; \\
  High Energy Physics Division, Argonne National Laboratory,
  9700 S.~Cass Ave., Argonne, IL, USA. \\
  E-mail:
  \email{aupadhye@anl.gov}}

\date{today}

\preprint{}

\abstract{
A scalar field dark energy candidate could couple to ordinary matter and photons, enabling its detection in laboratory experiments.  Here we study the quantum properties of the chameleon field, one such dark energy candidate, in an ``afterglow'' experiment designed to produce, trap, and detect chameleon particles.  In particular, we investigate the possible fragmentation of a beam of chameleon particles into multiple particle states due to the highly non-linear interaction terms in the chameleon Lagrangian.
Fragmentation could weaken the constraints of an afterglow experiment by reducing the energy of the regenerated photons, but this energy reduction also provides a unique signature which could be detected by a properly-designed experiment.  We show that constraints from the CHASE experiment are essentially unaffected by fragmentation for $\phi^4$ and $1/\phi$ potentials, but are weakened for steeper potentials, and we discuss possible future afterglow experiments.
}

\maketitle

\begin{document}

\section{Introduction}

Even as evidence for the cosmic acceleration continues to mount~\cite{Suzuki_etal_2012,Ade_etal_2013xvi,Anderson_etal_2013}, its cause remains a mystery.  The simplest dynamical explanation for this acceleration is a scalar field whose vacuum expectation value (VEV) corresponds to a small but nonzero potential~\cite{Ratra_Peebles_1988,Peebles_Ratra_1988,Reuter_Wetterich_1987,Wetterich_1988}.  Such a scalar could be a low-energy effective field associated with string theory or a modification to gravity.  In the absence of a symmetry forbidding such couplings, the scalar is expected to couple to Standard Model particles, mediating effects including fifth forces and oscillation.  Since these effects have not been observed, currently viable models include a non-linear ``screening mechanism'' by which the scalar interactions are suppressed in high-density environments.  Known screened models include: chameleons, which become effectively massive at high densities, reducing the range of their fifth force~\cite{Khoury_Weltman_2004a,Khoury_Weltman_2004b,Brax_etal_2004}; dilatons \cite{Brax:2010gi} and  symmetrons, which decouple from matter through the Damour-Polyakov mechanism \cite{Damour:1994zq} and a symmetry-restoring phase transition~\cite{Pietroni:2005pv,Olive_Pospelov_2007,Hinterbichler_Khoury_2010} in the symmetron case; and Galileons, whose non-canonical kinetic terms effectively decouple them from matter at high densities~\cite{Nicolis_Rattazzi_Trincherini_2008}.

Chameleon models have been shown to evade fifth force constraints in the laboratory~\cite{Kapner_etal_2007,Adelberger_etal_2007,Mota_Shaw_2006,Mota_Shaw_2007,Upadhye_Gubser_Khoury_2006,Brax_etal_2007c,Brax_etal_2009,Nesvizhevsky_etal_2002,Brax_Burrage_2011,Brax_Pignol_2011,Upadhye_2012,Brax_Pignol_Roulier_2013}, the solar system~\cite{Gubser_Khoury_2004,Adelberger_Heckel_Nelson_2003,Hu_Sawicki_2007}, compact astronomical objects~\cite{Upadhye_Hu_2009,Babichev_Langlois_2009,Upadhye_Steffen_2013,Brax_Davis_2013}, and the universe~\cite{Oyaizu_2008,Oyaizu_Lima_Hu_2008,Schmidt_etal_2009,Bernardeau_Brax_2011,Cabre_etal_2012,Jain_Vikram_Sakstein_2012,Brax_Valageas_2012,Li_etal_2012a,Lee_etal_2013,Vikram_etal_2013}.  Photon-coupled chameleon scalars, which could be produced through photon oscillation in much the same way as axions, have thus far escaped detection in laboratory ``afterglow'' experiments~\cite{Chou_etal_2009,Steffen_etal_2010,Brax_etal_2007b,Ahlers_etal_2008,Gies_Mota_Shaw_2008,Upadhye_Steffen_Weltman_2010,Upadhye_Steffen_Chou_2012} as well as astronomical probes~\cite{Burrage_Davis_Shaw_2009,Brax_Zioutas_2010,Brax_Lindner_Zioutas_2012}.  We are primarily interested in afterglow experiments, which attempt to produce chameleon particles through photon oscillation in a magnetic field.  Trapped by their matter interactions, these chameleon particles would oscillate back into photons even after the external photon source was switched off, leading to a photon afterglow by which they could be constrained.

On the other hand, questions have emerged regarding the stability of chameleon theories with respect to quantum corrections~\cite{Fujii_1997,Brax_Martin_2007,Hui_Nicolis_2010,Hinterbichler_Khoury_Nastase_2011}.  Ref.~\cite{Upadhye_Hu_Khoury_2012} showed that upcoming laboratory bounds will soon detect or exclude all chameleons with small $1$-loop corrections mediating gravitation-strength fifth forces at laboratory densities.  Since quantum effects cause chameleons to conflict with the predictions of Big Bang Nucleosynthesis~\cite{Erickcek_etal_2013,Erickcek:2013dea}, viable effective chameleon field theories must have cutoffs well below BBN energies of $\sim 1$~MeV.  Experiments seeking to test low-energy effective models such as chameleon dark energy must consider these quantum effects.

Here we discuss another such quantum effect, the production by ``fragmentation'' of many chameleon particles from fewer, higher-energy particles, in an afterglow experiment such as CHASE~\cite{Steffen_etal_2010}.  Models with large fragmentation rates typically have matter couplings large enough to satisfy the quantum stability bounds of~\cite{Upadhye_Hu_Khoury_2012}, and fragmentation can be large in models whose cutoffs are far lower than the BBN scale.  Although fragmentation is not predicted to be significant in CHASE for typical models, as we will show, it could provide a distinct signature in upcoming experiments. Photons sent into an afterglow experiment at one energy, after oscillation into chameleons which fragment, could emerge at lower energies in a predictable way.

Previous work~\cite{Upadhye_Steffen_Chou_2012} attempted to quantify fragmentation by considering a single two-body scattering event, with only limited success.  Chameleons in an afterglow experiment exist not as isolated particles, but as coherent states, which could fragment through their own momentum dispersion, through interactions with other wave packets, or through collision with large chameleon sources such as the chamber walls.  Working with coherent chameleon states, we show here that fragmentation due to interactions of two wavepackets is the dominant contribution.  Nevertheless, the nearly classical nature of these coherent states means that when the chameleon field is not substantially perturbed from its VEV, fragmentation is suppressed.  We then find that the fragmentation rate is unimportant over the parameter space excluded by CHASE for the most commonly-considered potentials, $\Lambda^4 \exp(\Lambda/\phi) \approx \Lambda^4 + \Lambda^5\phi^{-1}$ and $\lambda \phi^4/4!$.  However, there are regions in parameter space where fragmentation is large for steeper potentials.  Low-energy photons regenerated from fragmentation products would have evaded detection by CHASE, whose photomultiplier tube (PMT) detector was insensitive to energies below $\sim 1$~eV, but could potentially be detected by upcoming experiments.

The paper is organized as follows.  Section~\ref{sec:semi-classical_propagation_of_photons_and_chameleons} describes the propagation of a coherent chameleon wave packet in an afterglow experiment, as well as its oscillation to photons in an external magnetic field.  In Sec.~\ref{sec:fragmentation} we estimate the fragmentation rate due to momentum dispersion in a coherent state or in the interaction of two such states.  Section~\ref{sec:orders_of_magnitude} applies these results to afterglow experiments including CHASE, and Sec.~\ref{sec:conclusion} concludes.

\section{Semi Classical Propagation of Photons and Chameleons}%
\label{sec:semi-classical_propagation_of_photons_and_chameleons}

\subsection{Chameleons}

Chameleons have been introduced to model the late time acceleration of the expansion of the Universe~\cite{Brax_etal_2004} using a scalar field  whose dynamics are governed by a  potential $V(\phi)$ which depends on a single scale $\Lambda$
\begin{equation}
V(\phi)= \Lambda^4 f(\phi/\Lambda)
\end{equation}
where $\Lambda$ is determined by the present value of the dark energy, $\Lambda^4= 3 \Omega_{\Lambda 0} H_0^2 m_{\rm Pl}^2$ where $H_0$ is the Hubble rate now, i.e. $\Lambda \sim 2.4\times 10^{-12}\ {\rm GeV}$.
Hence we require that when $\phi\gg \Lambda$, $f\to 1$ so that the dynamics mimic the presence of an effective cosmological constant given by $\Lambda$. Moreover, $f$ is assumed to be decreasing and convex such that the second derivative of $V$ is positive guaranteeing  that the mass of the scalar field (in the absence of matter) is positive.
The original chameleon corresponds to the choice~\cite{Brax_etal_2004}
\begin{equation}
V(\phi)= \Lambda^4 \exp ((\frac{\Lambda}{\phi})^n)
\end{equation}
where $n>0$. When $\phi\gg\Lambda$, this behaves like a Ratra-Peebles model~\cite{Ratra_Peebles_1988,Peebles_Ratra_1988}
\begin{equation}
V(\phi)=\Lambda^4 + \frac{\Lambda^{4+n}}{\phi^n}+\dots
\end{equation}
where only the relevant terms have been kept. For such a model, dark energy is realised when $\phi \gg \Lambda$ which corresponds to a mass of the scalar field less than $\Lambda$, and therefore a range larger (and in practice much larger) than 1 mm where local tests of gravity are very stringent.  Hence this model of dark energy leads to the existence of a long range scalar force. Fortunately, this force can be screened in the solar system when the chameleon couples to matter.
Indeed the presence of matter has a direct effect on the potential which becomes
\begin{equation}
V_{\rm eff}(\phi)= V(\phi) +\frac{\beta}{m_{\rm Pl}} \phi.
\label{e:Veff}
\end{equation}
This effective potential is drastically different from $V(\phi)$ as it possesses a density-dependent minimum $\phi (\rho)$ with a mass $m(\rho)$ which increases with the density of matter.
This explains why chameleons cannot be seen in the solar system as large a body develops a thin shell which reduces the scalar field gradient in its vicinity.

Chameleons are coupled to matter via the rescaled metric $\tilde g_{\mu\nu}= e^{2\beta \phi/m_{\rm Pl}} g_{\mu\nu}$. As such this implies that chameleons are not coupled to photons at the classical level, though a photon coupling could of course be added to the action. At the quantum level,
the coupling of chameleons to fermions leads to a coupling whose origin follows from the non-conformal invariance of the fermionic measure in the path integral~\cite{Brax_etal_2009,Brax_etal_2011} (but see also ~\cite{Fujii_1997,Hui_Nicolis_2010}). In the following we shall take this coupling as
\be
S_\gamma= -\int d^4 x \sqrt{-g} \frac{\phi}{4M_\gamma}F^2
\ee
where
$M_\gamma$ is a coupling scale which is not fixed by the model. In the following, we will use $M_\gamma$ (or, equivalently, $\bgam = \Mpl/M_\gamma$) as a free phenomenological parameter which is constrained by experiments such as CHASE.

\subsection{Field equations}
The propagation and coupling between photons and chameleons is well documented. Chameleons can be produced by the Primakoff effect whereby photons of energy $k$ interact with a static magnetic field $B$ to create a chameleon particle. We will focus on experiments such as CHASE where a laser beam interacts with a magnetic field. In such situations, the photon beam can be considered to be in a quantum coherent state.  Due to the large occupation number, or the large flux, of photons which form such a coherent state, one is entitled to treat the incoming photons in a semi-classical way. This implies that the photon wave packet obeys the linear Maxwell equations coupled to the chameleon field which is also created  as a coherent field. As such, after linearising the Klein-Gordon equation around a background value $\phi_0$ associated with the minimum of the effective potential in the gas where photons propagate, the photon-chameleon system can be treated as a two level quantum mechanical problem with a transition probability from one state (the photon) to the other one (the chameleon). This approximation is valid as long as the non-linearities of the chameleon potential can be neglected and the chameleon-photon system remains coherent.
Here, we will revisit all this.

Let us expand the effective chameleon potential
\be
V_{\rm eff}(\phi) = \Lambda^4 + \frac{\Lambda^{4+n}}{\phi^n} +\beta \frac{\phi}{m_{\rm Pl}}
\ee
around the minimum $\phi_0$
\be
V_{\rm eff}(\phi_0+\delta\phi)= V_{\rm eff} (\phi_0) +\phi_0^4 \sum_{q>1} c_q (\frac{\Lambda}{\phi_0})^{n+4}(\frac{\delta \phi}{\phi_0})^q
\label{e:taylor_expansion}
\ee
where $c_q= (-1)^q \frac{n(n+1)\dots (n+q-1)}{q!}$.
We will identify the mass of the chameleon as $m_0^2= \frac{d^2 V_{\rm eff}}{d\phi^2}\vert_{\phi_0}$ and the self-coupling $\lambda= c_4 (\frac{\Lambda}{\phi_0})^{n+4}$.
The perturbation expansion is valid when all the coefficients of the terms in the series (\ref{e:taylor_expansion}) of order $p>2$ are small. At leading order this requires that $\phi_0\gtrsim  \Lambda$ and
for higher order terms $\delta\phi \lesssim \phi_0$. We will see how all the interaction terms in $\delta\phi^q$ in this effective potential affect chameleon wavepackets and can lead to the fragmentation of chameleons into less
energetic ones. The fragmentation rate is sensitive to how large the VEV $\phi_0$ is, how low the energies are compared to $\Lambda$ and how small the classical deviation $\delta\phi$ is with respect to $\phi_0$.
The fragmentation rate can be small even when $\phi_0\lesssim \Lambda$ provided $\delta\phi$ is small enough. In this case, no fragmentation occurs although the validity of the effective potential must be questioned as quantum corrections can be large. In the following, we shall always work in the perturbative regime.
The Klein-Gordon equation reads
\be
\partial^2\delta \phi  -\frac{\partial V_{\rm eff}}{\partial\phi}(\phi_0+\delta \phi)= -\frac{B}{M_\gamma} \partial_z A_y
\ee
where $\partial^2 = -\partial_t^2 +\partial_i^2$.  This is complemented with
the Maxwell equation which reads
\be
\partial^2 A_y=\frac{B}{M_\gamma} \partial_z \phi
\ee
where the magnetic field $B$  is in the $x$ direction and the photons propagate along the  $z$ direction.

As we are dealing with a quantum problem, these classical equations are replaced by operator-valued equations in the Heisenberg picture:
\be
\partial^2 \hat A_y=\frac{B}{M_\gamma} \partial_z  \delta \hat \phi
\ee
and the Klein-Gordon equation
\be
\partial^2\delta \hat\phi  -:\frac{\partial V_{\rm eff}}{\partial\phi}(\phi_0+\delta\hat \phi):= -\frac{B}{M_\gamma} \partial_z \hat A_y
\ee
where the canonical commutation relations are imposed too.
The equations of motion are normal ordered in order to remove tadpole singularities and to comply with the fact that the interaction Hamiltonian of the system is normal ordered~\cite{Weinberg_1995}.
We will analyse the time evolution of an initial photon coherent state and its mixing with the chameleon field.

\subsection{Coherent states and semi-classical treatment}

We will tackle the photon-chameleon mixing in the canonical formalism and we thus  expand the fields in creation and annihilation operators $a^\phi$ and $a^\gamma$
\be
 \delta \hat \phi= \int \frac{d^3k}{\sqrt{2\omega_\phi}}( e^{ik.x} a_{k}^\phi(t) + e^{-ik.x} a_{k}^{\dagger \phi}(t))
\ee
where $\omega_\phi(k)^2= k^2 +m_0^2$.
Similarly we have for the photons
\be
\hat A_y= \int \frac{d^3k}{\sqrt{2k}}( e^{ik.x} a_{k}^\gamma(t) + e^{-ik.x} a_{k}^{\dagger \gamma}(t))
\ee
where the normalisation of the creation and annihilation operators will be discussed later, in particular we shall see its relation with the coherence of the photon-chameleon system for relativistic chameleons.
These operators define a Fock space ${\cal H}_t$ at each time $t$ which will be unitarily related as long as the system remains coherent.
We  decompose $\hat A_y$ and $\delta\hat \phi$ into positive and negative frequency modes
$
\delta\hat\phi=\delta \hat \phi_+  +\delta\hat\phi_-,\ \hat A_y= \hat A_+ +\hat A_-
$
where
\be
\delta \hat \phi_+= \int\frac{ d^3k}{\sqrt{2\omega_\phi}} e^{ik.x} a_{k}^\phi(t), \hat A_+= \int \frac{d^3k}{\sqrt{2k}} e^{ik.x} a_{k}^\gamma(t)
\ee
which satisfy nice properties when acting on coherent states defined  as
\be
\vert A_k>= \exp{(-\frac{1}{2} \int d^3k \vert A_k(t)\vert^2)} e^{\int d^3 k A_k(t) a_k^{\dagger\gamma}(t)}\vert 0>
\ee
and
\be
\vert \phi_k>= \exp{(-\frac{1}{2} \int d^3k \vert \phi_k\vert^2)} e^{\int d^3 k \phi_k(t) a_k^{\dagger\phi}(t)}\vert 0>.
\ee
Indeed these coherent states are eigenmodes of the annihilation operators
$
a_k^\phi(t) \vert \phi_k>= \phi_k(t) \vert \phi_k>, \ a_k^\gamma(t) \vert A_k>= A_k(t) \vert A_k>.
$
This implies that the positive frequency parts of the operators satisfy
$
\delta\hat\phi_+\vert \phi_k>= \delta \phi_+, \ <\phi_k\vert \delta\hat\phi_-= <\phi_k\vert \delta \phi_-
$
where we have defined the classical fields as
\be
\delta \phi_+=\int \frac{d^3k}{\sqrt{2\omega_\phi}}\phi_k(t) e^{ik.x} ,\ \delta \phi_-=\int \frac{d^3k}{\sqrt{2\omega_\phi}} \phi_k^*(t) e^{-ik.x}
\ee
and
\be
A_+=\int \frac{d^3k}{\sqrt{2k}} A_k(t) e^{ik.x}  ,\ A_-=\int \frac{d^3k}{\sqrt{2k}} A_k^*(t) e^{-ik.x}.
\ee
As expected for coherent states, the averaged fields coincide with their classical values
$
<\phi_k\vert \delta \hat \phi\vert \phi_k>= \delta\phi_+ +\delta\phi_-\equiv \delta\phi
$
and
$
<A_k\vert \hat A_y\vert A_k>= A_+ +A_-\equiv A_y
$
Let us come back to the Klein-Gordon equation which reads now in terms of the non-linear interaction potential
\be
\partial^2\delta \hat\phi  -\phi_0^3 \sum_{q>1} qc_q (\frac{\Lambda}{\phi_0})^{n+4}:(\frac{\delta \hat \phi}{\phi_0})^{q-1}:= -\frac{B}{M_\gamma} \partial_z \hat A_y.
\ee
This is a non linear equation due to the terms of order $q>2$. Let us focus on the non-linear terms first.
As we have normal ordered these terms, we have
$
:(\delta \hat \phi)^{q-1}:= \sum_{i=0}^{q-1} C_{q-1}^i \delta \hat \phi_-^i \delta \hat \phi_+^{q-1-i}
$
which implies that the quantum averaged value satisfies
$
<\phi_k\vert :(\delta \hat \phi)^{q-1}:\vert \phi_k>= (\delta \phi)^{q-1}
$
This is a fundamental property which allows us to write the averaged value of the Klein-Gordon equation in the state $\vert \phi_k>$
\be
\partial^2\delta \phi  -\phi_0^3 \sum_{q>1} qc_q (\frac{\Lambda}{\phi_0})^{n+4}(\frac{\delta \phi}{\phi_0})^{q-1}= -\frac{B}{M_\gamma} \partial_z  A_y
\ee
which is nothing but the Klein-Gordon equation for the classical fields.
This result is true as long as one can neglect the time dependence of the coherent states $\vert A_k>$ and $\vert \phi_k>$.
We will see that this result holds as long as time $t$ is less than the coherence time $t_{\rm coh}={k/4m_0^2}$.

\subsection{Time evolution of the quantum operators}

In the following we will first focus on situation where $\delta\phi/\phi_0\lesssim  1$ implying that the classical Klein-Gordon equation can be linearised
\be
\partial^2\delta \phi  -m_0^2 \delta\phi= -\frac{B}{M_\gamma} \partial A_y.
\ee
Together with Maxwell's equation
\be
\partial^2 A_y=\frac{B}{M_\gamma} \partial_z \phi
\ee
thsi leads to the time evolution of the quantum operators.

The system of linear equations can be diagonalised easily by introducing
the vectors
\be
v_k(t)=\left (\begin{array}{c}
A_k(t)\\
i\delta\phi_k (t)
\end{array}\right )
\ee
such that the mode equations become
\be
-\partial_t^2 v_k =U_k v_k
\ee
where the evolution is defined by the matrix
\begin{equation}
U_k=
\left(\begin{array}{cc}
k^2& -\frac{B k}{M_\gamma} \\
-\frac{B k}{M_\gamma} & k^2 +m_0^2
\end{array}\right).
\end{equation}
One can diagonalise this evolution matrix as
$
U_k= P^T D P
$
where the mixing matrix reads
\be
P=\left(\begin{array}{cc}
\cos\theta& -\sin\theta \\
\sin\theta & \cos\theta
\end{array}\right)
\end{equation}
and the mixing angle is defined as
\be
\tan 2 \theta =\frac{2Bk}{M_\gamma m_0^2}.
\ee
The eigenfrequencies are defined by the diagonal matrix
\be
D=\left(\begin{array}{cc}
\omega_-^2& 0 \\
0 & \omega_+^2
\end{array}\right)
\end{equation}
corresponding to the eigenmodes of the system
\be
\omega_\pm^2= k^2 +m^2\frac{\cos 2\theta \pm 1}{2 \cos 2\theta}.
\ee
The associated eigenvectors are identified with
$
u_k(t)=Pv_k(t)
$
from which we can select the right moving solutions
$
u_k(t)= E(t) u_k(0)
$
where the evolution operator is given by
\be
E(t)=\left(\begin{array}{cc}
e^{-i\omega_-^2 t}& 0 \\
0 & e^{-i\omega_+^2 t}
\end{array}\right)
\end{equation}
and therefore the classical solutions are such that
\be
v_k (t)=P^TE(t)P v_k(0)
\ee
where initially there is no mixing between the photons and the chameleons and the fields are canonically normalised
\be
v_k(0)=\left (\begin{array}{c}
{A_k}\\
{i\phi_k}
\end{array}\right ),
\ee
here $A_k$ and $\phi_k$ are the initial Fourier modes of the photon and chameleon waves.
This implies that the right moving solutions evolve according to
\be
\left (\begin{array}{c}
A_+(t)\\
i\delta \phi_+(t)
\end{array}\right )= P^T E(t) P \left (\begin{array}{c}
A_+(0)\\
i\delta \phi_+(0)
\end{array}\right )
\ee
which corresponds to the equality between coherent states
\be
\left (\begin{array}{c}
\hat A_+(t)\\
i\delta\hat\phi_+(t)
\end{array}\right )\vert A_k>\otimes \vert \phi_k>= P^T E(t) P \left (\begin{array}{c}
\hat A_+(0)\\
i\delta\hat \phi_+(0)
\end{array}\right )\vert A_k>\otimes \vert \phi_k>
\ee
where we have used the fact that the coherent states have a negligible time evolution for $t$ less than the coherence time.
Therefore we find that the evolution of the quantum operators in the Heisenberg picture is given by
\be
\left (\begin{array}{c}
\hat A_+(t)\\
i\delta\hat\phi_+(t)
\end{array}\right )= P^T E(t) P \left (\begin{array}{c}
\hat A_+(0)\\
i\delta\hat \phi_+(0)
\end{array}\right )
\ee
together with their complex conjugates
\be
\left (\begin{array}{c}
\hat A_-(t)\\
-i\delta\hat\phi_-(t)
\end{array}\right )= P^T E^\dagger (t) P \left (\begin{array}{c}
\hat A_-(0)\\
-i\delta\hat \phi_-(0)
\end{array}\right ).
\ee
We have introduced the initial operators
\be
\delta \hat \phi_+(0)= \int \frac{d^3k}{\sqrt{2\omega_\phi}} e^{ik.x} a_{k}^\phi
\ee
 and
\be
\hat A_+(0)= \int \frac{d^3k}{\sqrt{2k}} e^{ik.x} a_{k}^\gamma
\ee
defined by the annihilation operators $a_k^{\gamma,\phi}$.
Using all these results we obtain that the annihilation operators evolve according to
\be
\frac{a_k^\gamma(t)}{\sqrt{2k}}=(\cos^2 \theta e^{-i\omega_- t} +\sin^2\theta e^{-i\omega_+ t})\frac{a_k^\gamma}{\sqrt{2k}} + \frac{\sin 2\theta}{2}(e^{-i\omega_+ t}-e^{-i\omega_- t})\frac{ia^{\phi}}{\sqrt{2\omega_\phi}}
\ee
and
\be
\frac{ia_k^\phi(t)}{\sqrt{2\omega_\phi}(k)}=(\cos^2 \theta e^{-i\omega_+ t} +\sin^2\theta e^{-i\omega_- t})\frac{ia_k^\phi}{\sqrt{2\omega_\phi}(k)} + \frac{\sin 2\theta}{2}(e^{-i\omega_+ t}-e^{-i\omega_- t})\frac{a^{\gamma}}{\sqrt{2k}}
\ee
for the evolution of the operators and
\be
\frac{A_k(t)}{\sqrt{2k}}=(\cos^2 \theta e^{-i\omega_- t} +\sin^2\theta e^{-i\omega_+ t})\frac{A_k}{\sqrt{2k}} + \frac{\sin 2\theta}{2}(e^{-i\omega_+ t}-e^{-i\omega_- t})\frac{i\phi_k}{\sqrt{2\omega_\phi (k)}}
\ee
\be
\frac{i\phi_k(t)}{\sqrt{2\omega_\phi (k)}}=(\cos^2 \theta e^{-i\omega_+ t} +\sin^2\theta e^{-i\omega_- t})\frac{i\phi_k}{\sqrt{2\omega_\phi (k)}} + \frac{\sin 2\theta}{2}(e^{-i\omega_+ t}-e^{-i\omega_- t})\frac{A_k}{\sqrt{2k}}
\ee
for the amplitudes.
We can now check  that the equal time commutation relations
\be
[a_k^{\gamma,\phi}(t),a_{k'}^{\gamma,\phi}(t)]=\delta^{(3)}(k-k')
\ee
are only satisfied when $k\gg m_0$. In this case, all the Hilbert spaces ${\cal H}_t$ are unitarily equivalent.

To leading order, we find that when the mixing is small
\be
\omega_+^2 -\omega_-^2= \frac{m_0^2}{\cos 2 \theta}
\ee
and therefore
\be
\omega_+\approx\omega_-+ \frac{m_0^2}{2\omega_- \cos 2 \theta}.
\ee
Starting from no chameleon initially, we find that
\be
\phi_k(t) \approx ({\sin 2\theta}\sin\frac{t}{t_{coh}}) e^{-i\omega_- t}{A_k}
\ee
when $k\gg m_0$.
This gives the usual transition probability from one photon state to a chameleon state
\be
P_{\gamma\to\phi}(t)={\sin^2 2\theta}\sin^2\frac{t}{t_{coh}}
\ee
with $t_{coh}=\frac{4\omega_-\cos 2 \theta}{m^2_0}$.
Notice that, to leading order, $A_k(t) a_k^{\dagger\gamma}(t) $ is time independent for $t\lesssim t_{coh}$ implying that our assumption was justified and the coherent states remain coherent all this time.

\section{Fragmentation}
\label{sec:fragmentation}

\subsection{Fragmentation to lowest order}

The evolution of the operators and the states that we have considered so far correspond, in the interaction picture, to the states and the operators evolving with the part of the
Hamiltonian which does not include the interaction terms. In fact the evolution operator in the interaction picture can be written as
\be
U(t)=\exp{(- i\int_{t_0}^{t_1} dt H_{\rm int})}
\ee
where the interaction Hamiltonian is
\be
H_{\rm int}=  \phi_0^4 \sum_{p>2} c_p (\frac{\Lambda}{\phi_0})^{n+4}\int d^3 x :(\frac{\delta \hat \phi}{\phi_0})^q:.
\ee
To leading order we have
\be
U(t)= 1-i \int_{t_0}^{t_1} dt H_{\rm int}
\ee
where we shall focus on one particular interaction term
\be
H_q=  c_q \phi_0^4   (\frac{\Lambda}{\phi_0})^{n+4}\int d^3 x:(\frac{\delta \hat \phi(x,t)}{\phi_0})^q:\ .
\ee
We are interested in the transition probability between the initial chameleon-photon  state $\vert A_k(t_0)>\otimes\vert \phi_k(t_0)>$ and the final chameleon-photon  state where one free chameleon has been created with momentum $k_1$, $\vert A_k (t_1)>\otimes a^{\dagger \phi}_{k_1}(t_1)\vert \phi_k(t_1)>$.
To leading order we can omit the photon part of the state and consider the transition between $\vert \phi_k(t_0)>$ and $a^{\dagger \phi}_{k_1}(t_1)\vert \phi_k(t_1)>$
where $w_+\approx w_\phi$.
Let us first evaluate the matrix element of the interaction part of the Hamiltonian
\be
<\phi_k(t_1)\vert a_{k_1}^\phi  :\hat \phi^q(x,t):\vert \phi_k(t_0)>\approx\sum_{i=0}^q C^j_q <\phi_k(t_0)\vert a_{k_1}^\phi \hat \phi_-^j(x,t) \hat\phi_+^{q-j}(x,t)\vert \phi_k(t_0)>
\ee
where we have neglected the time dependence of the states and to leading order
\be
\hat \phi_+(x,t)=\int \frac{d^3k}{\sqrt{2 \omega_\phi}(k)} e^{-i \omega_\phi(k) t +ik.x} a_k^\phi.
\ee
Using the commutation relation $[a_{k_1}^\phi,a^{\dagger\phi}_{k_2}]=\delta^{(3)}(k_1-k_2)$, and
\be
[a_{k_1}^\phi,\hat \phi_- (x,t)]= \frac{e^{i \omega_\phi(k_1) t -ik_1.x}}{\sqrt{2 \omega_\phi(k_1)}}
\ee
we find that
\ba
&&< \phi_k (t_1)\vert a_{k_1}^\phi(t_1) \hat \phi_-^j(x,t) \hat\phi_+^{q-j}(x,t)\vert \phi_k(t_0)>\nonumber \\ &\approx& j \phi_-(x,t)^{j-1} \phi_+(x,t)^{q-j}\frac{e^{i\omega_\phi(k_1) (t-t_1)-ik_1.x}}{\sqrt{2 \omega_\phi (k_1)}}+\phi_{k_1}(t_1) \phi_-(x,t)^{j} \phi_+(x,t)^{q-j}\nonumber \\
\ea
and therefore
\ba
&&<\phi_k(t_1)\vert  a_{k_1}^\phi(t_1):\hat \phi^q(x,t) :\vert \phi_k(t_0)>\nonumber \\ &&\approx \sum_{j=0}^q C^j_q j \phi_-(x,t)^{j-1} \phi_+(x,t)^{q-j}\frac{e^{i\omega_\phi(k_1) (t-t_1) -ik_1.x}}{\sqrt{2 \omega_\phi(k_1)}}+\phi_{k_1}(t_1) \phi_-(x,t)^{j} \phi_+(x,t)^{q-j})\nonumber \\
\ea
leading to
\be
<\phi_k(t_1)\vert a_{k_1}^\phi(t_1) :\hat \phi^q(x,t):\vert \phi_k(t_0)>
\approx q(\delta \phi(x,t))^{q-1} \frac{e^{i\omega_\phi(k_1) (t-t_1)-ik_1.x}}{\sqrt{2 \omega_\phi(k_1)}} +\phi_{k_1}(t_1)(\delta \phi(x,t))^{q}.
\ee
We have consistently assumed that the states evolve slowly, which requires that $\vert t_1 -t_0\vert \ll t_{coh}$.
As a result we have, to leading order,
\ba
&&< \phi_k (t_1)\vert  a_{k_1}^\phi(t_1) U \vert \phi_k (t_0)> \approx \phi_{k_1}(t_1)(1-ic_q  \phi_0^4  (\frac{\Lambda}{\phi_0})^{n+4}\int d^3 x dt(\frac{\delta  \phi(x,t)}{\phi_0})^q) \nonumber \\ &&
-i qc_q  \phi_0^3  (\frac{\Lambda}{\phi_0})^{n+4}\int d^3 x dt(\frac{\delta  \phi(x,t)}{\phi_0})^{q-1} \frac{e^{i\omega^\phi (k_1) (t-t_1)-ik_1.x}}{\sqrt{2 \omega_\phi (k_1)}}.\nonumber \\
\ea
The first term corresponds to transition probability from one photon to one chameleon renormalised by the presence of the interaction term. The second term is the result of the non-linear interaction in the potential.

\subsection{Monochromatic chameleons}

To go further, we
now focus on monochromatic photon beams in the initial state. In this case,
\be
A_k= \sqrt{2k_\star} A_\gamma \delta^{(3)}(k-k_\star)
\ee
where $k_\star$  is the energy of the beam.
Similarly we have
\be
\phi_k (t)= {\sin 2\theta}\sin(\frac{t}{t_{coh}}) e^{-ik t}A_\gamma \delta^{(3)}(k-k_\star)
\ee
and therefore
\be
\delta \phi (x,t)= {\sin 2\theta}\sin(\frac{t}{t_{coh}}) e^{ik_\star.x -ik_\star t}A_\gamma
\ee
for a plane wave representing the time evolution of the classical chameleon field. In this case the renormalisation contribution in $\int d^3x dt (\delta\phi (x,t))^q$ vanishes. Subtracting the contribution from
the free evolution of the system, we  have the part of  the matrix element
\ba
&&
\int d^3 x dt(\frac{\delta  \phi(x,t)}{\phi_0})^{q-1} \frac{e^{i\omega_\phi (k_1) (t-t_1)-ik_1.x}}{\sqrt{2 \omega_\phi (k_1)}}\nonumber \\
&&\approx  \frac{e^{-i\omega_\phi (k_1)  t_1}}{\sqrt{2 \omega_\phi (k_1)} }(\frac{\delta\phi}{\phi_0})^{q-1} (2\pi)^4 \delta^{(3)}(k_1-(q-1)k_\star) \delta (\omega_\phi (k_1) - (q-1) \omega_\phi (k_\star))\nonumber
\ea
where we have taken into account that the variation of the exponential is much faster than the one of $\sin(\frac{t}{t_{\rm coh}})$ when $k\gg m_0$ and we have
\be
\delta\phi={\sin 2\theta}\sin(\frac{t_\star}{t_{\rm coh}})A_\gamma
\ee
where $t_\star$ is a typical time between $t_0$ and $t_1$.
This result simply expresses that the chameleon can only be created with an impulsion $(q-1)k_\star$. This can be understood from a Feynman diagram point of view as saying that one chameleon has been created from
$(q-1)$ photons extracted from the coherent state all with a momentum $k_\star$.

We can generalise this result to changes in the particle number greater than one.
Consider the  fragmentation process happening  when the state $(\prod_{i=1}^{p} a^{\dagger \phi}_{k_i}(t_1))\vert \phi_k(t_1)>$
is created thanks to the operator $:\hat \phi^q (x,t):$. In this case we obtain a matrix element of the form
\begin{eqnarray}
-i a_{q,p}c_q \phi_0^{4-p}
&\!&
\!\!\!
(\frac{\Lambda}{\phi_0})^{n+4}(\prod_{i=1}^{p} \frac{e^{-i\omega_\phi (k_i) t_1}}{\sqrt{2 \omega_\phi (k_i)}}) (\frac{\delta\phi}{\phi_0})^{q-p} (2\pi)^4 \delta^{(3)}(\sum_{i=1}^{p}k_i-(q-p)k_\star)
\nonumber\\
&\!&
\times\,
 \delta (\sum_{i=1}^{p}\omega_\phi (k_i)- (q-p)\omega_\phi(k_\star))
\end{eqnarray}
where $a_{q,p}= q(q-1)\dots (q-p+1)$.
This expresses the fact that $p$ chameleons are created from the 4-momentum $(q-p)k_\star$.
Taking the square of this matrix element and integrating over the momenta $k_i$, we obtain a probability per unit time and unit volume
\begin{eqnarray}
\frac{dP}{V dt}
&=&
c_q^2 a_{q,p}^2 \phi_0^8 (\frac{\Lambda}{\phi_0})^{2n+8}(\frac{\delta\phi}{\phi_0})^{2(q-p)} \int \prod_{i=1}^{p} (\frac{d^3k_i}{2\omega_\phi (k_i)\phi_0^2} (2\pi)^4 \delta^{3}(\sum_{i=1}^{p}k_i-(q-p)k_\star)
\nonumber\\
&\,&
\times\,
\delta (\sum_{i=1}^{p}\omega_\phi (k_i) - (q-p)\omega_\phi(k_\star)).
\end{eqnarray}
This is very easily interpreted noticing that the coherent state provides a momentum $(q-p)k_\star$ which fragments into $p$ chameleon momenta. In the frame where the initial chameleon is at rest (which exists as the chameleon is massive) and $k_\star=(m_0,0,0,0)$, we see that changes in the particle number in a perfectly monochromatic state are possible only when $q\ge 2p$.  That is, chameleons in such a state can only be destroyed, not produced.  Nevertheless, the rates of such processes is  instructive as the created chameleons are free particles emerging from the initial coherent state.
The phase space integral can be estimated as
\be
V^{-1}\frac{dP}{ dt}(t_0)\sim  \phi_0^4 (\frac{\Lambda}{\phi_0})^{2n+8}(\frac{\delta\phi}{\phi_0})^{2(q-p)} (\frac{m_0}{\phi_0})^{2(p-2)}
\label{e:monochromatic}
\ee
which depends on $t_0$ via $t_\star$. This result must be averaged over $t_*$ in order to get the total probability per unit volume and time
of creating $p$ chameleons from the laser beam. As $t_*<< t_{\rm coh}$, averaging is equivalent to averaging $t_\star^{2(q-p)}$, i.e. introducing a factor of $1/(2(q-p)+1)$ and $t_\star=L$
in a cavity experiment where   the volume $V$ is the length $L$ times the beam section $S$.

Evidently from~(\ref{e:monochromatic}), processes which change the particle number in a chameleon beam are suppressed when:
\begin{enumerate}
\item the VEV is large, $|\phi_0| \gg \Lambda$, corresponding to the perturbative regime;
\item the oscillation amplitude is small, $|\delta\phi| \ll |\phi_0|$, implying fewer available chameleon particles; or
\item the available center-of-mass energy is small, $E \ll |\phi_0|$, limiting the phase space for such processes.
\end{enumerate}
We will see that these three conditions apply more generally to states with a nonzero momentum dispersion, as well as to interactions between two coherent states.


\subsection{Chameleon wave packet}
Let us start with some simplifying assumptions.  First, we  approximate the momentum scatter by assuming two equal chameleon populations of slightly different momenta.  Rather than a rest frame, there exists a centre-of-momentum (CM) frame in which the chameleon momenta are $\vec k_\pm = (E, 0, 0, \pm \sigma)^T$ with $E^2 = m_0^2 + \sigma^2$.  Second, we turn off the photon-chameleon oscillation after the chameleon amplitude has built up to some value $\delphi_0$, as obtained at the end of an optical cavity.

We then take
\begin{equation}
\phi_{\vec k}(t) = \sqrt{2E} \delphi_0 \delta(k_x) \delta(k_y)
\times \frac{1}{2}\left[\delta(k_z-\sigma) - \delta(k_z+\sigma)\right]
\end{equation}
as a first step corresponding to a wave packet comprising two monochromatic waves. More generally, a wave packet with a symmetrical distribution $g(k_z/\sigma)$ centred around 0 and with a width $\sigma$ can be obtained as
\begin{equation}
\phi_{\vec k}(t) =\int_0^\infty ds g(s) \sqrt{2E} \delphi_0 \delta(k_x) \delta(k_y)
\times \frac{1}{2}\left[\delta(k_z-s\sigma) - \delta(k_z+s\sigma)\right]
\end{equation}
with $E^2 = m_0^2 + s^2 \sigma^2$.
As the distribution functions $g$ are of order one and converge to zero at infinity, the simpler calculation with two $\delta$ functions gives us the right order of magnitude for the decay rate.

Then the fragmentation probability per unit volume and time becomes
\begin{eqnarray}
\frac{d\mathcal P}{Vdt}
&=&
c_q^2 a_{q,p}^2\phi_0^8
\left(\frac{\Lambda}{\phi_0}\right)^{2n+8} \!\!\!
\left(\frac{\delphi_0}{2\phi_0}\right)^{2(q-p)} \!\!\!\!\!
\int\prod_{i=1}^p \left(\!\frac{d^3 k_i}{2\omega_\phi (k_i)\phi_0^2}\!\right)
(2\pi)^4
\delta\!\!\left(\sum_{i=1}^p \omega_\phi (k_i) - (q-p)E\!\!\right)
\nonumber\\
&&
\times \quad
\delta\left(\sum_{i=1}^p k_{ix}\right)
\delta\left(\sum_{i=1}^p k_{iy}\right)
\sum_{j=0}^{q-p} C_{q-p}^j\delta\left((2j+p-q)\sigma-\sum_{i=1}^p k_{iz}\right).
\label{e:rate_general}
\end{eqnarray}
Each term represents the fragmentation into $p$ chameleons whose momenta along the $z$ axis vary between $(p-q)\sigma$ and $(q-p)\sigma$.

Such a process involves $q$ chameleons where $(q-p)$ emerge from the coherent states while $p$ free chameleons appear due to the interaction in the Hamiltonian $H_q$. The number of created chameleons is
\be
\Delta N_q= 2p-q
\ee
which can be very large. In the following, we will estimate the probability per unit volume and time for the two cases $\sigma \ll m_0$ (low momentun scatter) and $\sigma \gg m_0$ (high momentum scatter).

\subsubsection{Low momentum scatter}

First, consider the case $\sigma \ll m_0$. This is particularly appropriate for cavity experiments such as CHASE since $\sigma \sim 2 \times 10^{-5}$~eV and $m_0 \gg 10^{-5}$~eV in a chamber a few centimeters in radius. In this limit, $\omega_\phi (k_i) = m_0 + k_i^2/(2m_0)$, so the energy delta function becomes
\be
\delta\left(\sum_{i=1}^p |\vec k_i|^2 / (2m_0) + (2p-q)m_0 - (q-p)\sigma^2/(2m_0)\right).
\ee
This restricts the volume of integration to a $(3p-1)$ dimensional sphere whose radius squared is $(q-p)\sigma^2 - 2(2p-q)m_0^2$.  The $k_{ix}$ delta function further restricts the integration volume to a plane in $k_i$-space which passes through the origin, meaning that it lowers the dimensionality of the sphere by one without changing the radius; the same is true of the $k_{iy}$ delta function.  Meanwhile, the $k_{iz}$ delta function restricts integration to a plane which passes a distance $|2j+p-q|\sigma/\sqrt{p}$ from the origin.  The final result is a $(3p-4)$-dimensional sphere of radius
\begin{equation}
\kappa_j
=
\sqrt{(q-p)\sigma^2 - 2(2p-q)m_0^2 - \frac{(2j+p-q)^2}{p}\sigma^2},
\label{e:kappa_j}
\end{equation}
with the phase space integral vanishing for imaginary $\kappa_j$.

Clearly the most kinetic energy will be available for $j \approx (q-p)/2$, that is, equal numbers of left-moving and right-moving particles.  Then $\kappa_j$ will be real if $q-p \geq 2(2p-q)m_0^2/\sigma^2$.  Suppose that the increase in the total number of particles, $2p-q$, is just one.  For instance when  $\sigma = 2\times 10^{-5}$~eV and $m_0 = 2 \times 10^{-3}$~eV~$\sim \Lambda$ we have $q-p \geq 20000$ which corresponds to a very large number of chameleons.

Carrying out the integral in~(\ref{e:rate_general}), we find
\begin{eqnarray}
\frac{d\mathcal{P}}{Vdt}
&=&
c_q^2 a_{q,p}^2\phi_0^8
\left(\frac{\Lambda}{\phi_0}\right)^{2n+8} \!\!\!
\left(\frac{\delphi_0}{2\phi_0}\right)^{2(q-p)} \!
\frac{32\pi^{(3p+5)/2} m_0}{(2m_0\phi_0^2)^p \Gamma\left(\frac{3p-3}{2}\right)}
\sum_{j=0}^{q-p} C_{q-p}^j \kappa_j^{3p-5}
\nonumber\\
&=&
c_q^2 a_{p,q}^2\frac{\phi_0^8 m_0}{\kappa_j^5} \!\!
\left(\frac{\Lambda}{\phi_0}\right)^{2n+8} \!\!\!
\left(\frac{\delphi_0}{\phi_0}\right)^{2(q-p)} \!\!\!
\frac{32\pi^{5/2}}{\Gamma\left(\frac{3p-3}{2}\right)}
\sum_{j=0}^{q-p} \frac{C_{q-p}^j}{2^{2(q-p)}} \!\!
\left(\!\!\frac{\pi^{3/2}\kappa_j^3}{2m_0\phi_0^2}\!\right)^p.
\label{e:rate_small_sigma}
\end{eqnarray}
Since $p$ and $q-p$ can be very large, we must ensure that this expression does not diverge as $p,\, q \rightarrow \infty$.  This implies that $|\delphi_0 / \phi_0|$ must be sufficiently small.  On the other hand, $n$ is fixed, so $\phi_0$ can be smaller than $\Lambda$ without the fragmentation rate diverging. Of course, $\phi_0 \ll \Lambda$ implies large quantum corrections to the potential, hence corrections to the phase shift associated with chameleon reflection from walls, but afterglow experiments are insensitive to this~\cite{Upadhye_Steffen_Chou_2012}.

In order to put an upper bound on fragmentation due to nonzero $\sigma$, let us assume that $p$ and $q-p$ are large but $2p-q = 1$, so that exactly one new particle is created.  Since $p \approx q-p$, $\kappa_j \approx p^{1/2}\sigma$, and we have
\begin{equation}
\frac{d\mathcal{P}}{Vdt}
\lesssim
\frac{32\pi^{2}}{\sqrt{3}}
\frac{\phi_0^8 m_0}{\sigma^5} \!\!
\left(\!\frac{\Lambda}{\phi_0}\!\right)^{\!\!2n+8} \!\!\!
\left(\!\frac{\delphi_0}{\phi_0}\!\right)^{\!\!2p}
\frac{c_{2p}^2 a_{2p,p}^2}{p^3}
\left[\frac{2 e \pi \sigma^2}{3 (2 m_0 \phi_0^2)^{2/3}}\right]^{3p/2}.
\label{e:rate_p}
\end{equation}
The total fragmentation rate is obtained by summing over all $p$.  Assuming $\delphi_0\lesssim \phi_0$, this sum will converge if the quantity in square brackets is less than unity, that is, if
\begin{equation}
\sigma
\lesssim
\sqrt{\frac{3}{2e\pi}} \left(2m_0 \phi_0^2\right)^{1/3}
\sim
m_0^{1/3} \phi_0^{2/3}.
\label{e:sigma_convergence}
\end{equation}
   Furthermore, if $\delphi_0/\phi_0 \ll 1$, then the sum will converge for
\begin{equation}
\sigma \lesssim  \left|\delphi_0/\phi_0\right|^{-2/3} m_0^{1/3} \phi_0^{2/3}.
\label{e:convergence_small_dphi}
\end{equation}
This can be a considerably weaker condition than~(\ref{e:sigma_convergence}).

\subsubsection{High momentum scatter}

Next, consider the limit $\sigma \gg m_0$.  This could apply, for example, to a relativistic standing wave or to two separate chameleon pulses passing through one another.  In this case the integral in~(\ref{e:rate_general}) is more difficult to evaluate.  By dimensional analysis, we can estimate
\begin{eqnarray}
\frac{d\mathcal{P}}{Vdt}
&\sim&
c_q^2 \frac{\phi_0^8}{\sigma^4}
\left(\frac{\Lambda}{\phi_0}\right)^{2n+8}
\left(\frac{\delphi_0}{\phi_0}\right)^{2(q-p)}
\left(\frac{\sigma^2}{\phi_0^2}\right)^p.
\label{e:rate_large_sigma}
\end{eqnarray}
At large $p$ and $q-p$ the summation over such terms diverges unless $\delphi_0$ and $\sigma$ are sufficiently small.  Assuming, as above, that $p = q-p + 1 \gg 1$, the convergence criterion is approximately
\be
\left|\delphi_0 \sigma / \phi_0^2\right| < 1.
\label{e:convergence_large_sigma}
\ee
 Thus for $\delphi_0/\phi_0 \ll 1$, the fragmentation rate will be finite even for $\sigma$  larger than $\phi_0$.

\section{Estimates of Fragmentation in Experiments}
\label{sec:orders_of_magnitude}

\subsection{Afterglow experiments}

\begin{figure}[tb]
\begin{center}
\includegraphics[width=4in]{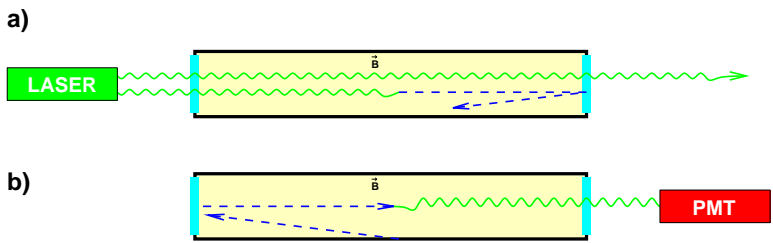}
\caption{A simple model of an afterglow experiment, from~\cite{Upadhye_Steffen_Chou_2012}.  $\left.a\right)$~Production phase.  Photons stream via entrance and exit windows through a vacuum chamber with magnetic field.  Some photons oscillate into chameleon particles, which reflect from the windows and are trapped inside the chamber.  $\left.b\right)$~Afterglow phase.  The photon source is turned off and an external PMT detector is uncovered.  The population of trapped chameleons regenerates photons through oscillation.  Some of these photons emerge from the exit window and reach the PMT. \label{f:gammev}}
\end{center}
\end{figure}

Consider a chameleon afterglow experiment such as the one shown in Fig.~\ref{f:gammev}.  Photons are streamed via entrance and exit windows through a vacuum chamber containing a magnetic field $B$.  Chameleon particles produced through photon oscillation in this magnetic field are trapped inside the chamber if the chameleon effective mass in the chamber walls exceeds the total energy of an individual chameleon particle inside the chamber.  Trapped chameleon particles regenerate photons through oscillation, implying a photon afterglow emitted by the chamber even after the external photon source has been switched off.  As shown in Fig.~\ref{f:gammev}, an external detector can be used to search for this afterglow, and, hence, to constrain the underlying models.

Fragmentation can weaken the constraints of an afterglow experiment by converting trapped high-energy chameleon particles into lower-energy particles whose regenerated photons are not energetic enough to be detected.  There are three different processes in which fragmentation could be significant:
\begin{enumerate}
\item propagation of a coherent chameleon state through the chamber;
\item reflection of such a state from the chamber wall;
\item interaction of two such states passing through one another.
\end{enumerate}
The first two processes are one-state processes which can be approximated using~(\ref{e:rate_small_sigma}) and bounded using~(\ref{e:rate_p}).  As an example, assume $|\delphi/\phi| = 10^{-5}$ for CHASE and estimate $m_0 \sim \phi_0 \sim \Lambda$.  Then the $p$-dependent factor in~(\ref{e:rate_p}) is $(4\times 10^{-16})^p c_{2p}^2a_{2p,p}^2 / p^3$.  Since the minimum $p$ for the net production of one chameleon particle is $\sim 10^4$, fragmentation within the coherent state is entirely negligible.  As we will show, $\delphi/\phi$ decreases when the chameleon state approaches a chamber wall, so that fragmentation during wall reflection should also be small.  In the next section we generalize this example, showing that the first two processes in the list above should not contribute significantly to chameleon fragmentation in CHASE-like experiments.  Then we proceed to estimate the third of these processes in afterglow experiments.

\subsection{One-state fragmentation}

Out of the quantities in (\ref{e:convergence_small_dphi},~\ref{e:convergence_large_sigma}), $\sigma$ is a parameter of the experiment while $\phi_0$ and $m_0$ are readily determined by minimizing the effective potential~(\ref{e:Veff}).  The chameleon amplitude $\delta\phi$ has a more complicated dependence on the particulars of the experiment.
We will determine the order of magnitude of $\delta\phi$ in experimentally relevant situations.

Let us a consider photons emitted by a powerful source of power $P_\gamma$ with a cross section $S$. The photon flux measured by the norm of the Poynting vector is $\Phi=P_\gamma/S$ where
\be
\Phi= k^2 A_\gamma^2
\ee
and $A_\gamma$ is the norm of the vector potential, implying that
\be
A_\gamma= \frac{\sqrt{\Phi}}{k}
\ee
corresponding to the amplitude of the photon wave packet. Assuming that the mixing between chameleons and photons is small, the mixing angle is
\be
\theta= \frac{kB}{M_\gamma m_0^2}.
\ee
The maximal amplitude of the chameleon wave packet is
\be
\delta\phi= 2\theta A_\gamma.
\ee
Using the fact that
\be
m_0^2=  n(n+1) \frac{\Lambda^{n+4}}{\phi_0^{n+2}}
\ee
we obtain
\be
\frac{\delta \phi}{\phi_0}= \frac{2\sqrt\Phi B}{n(n+1)M_\gamma \Lambda^{n+1}}\phi_0^{n+1}.
\ee
Upon using
\be
\phi_0^{n+1}= \frac{n m_{\rm Pl} \Lambda^{n+4}}{\beta \rho}
\ee
we get
\be
\frac{\delta\phi}{\phi_0}= \frac{\beta_\gamma}{\beta} \frac{2\sqrt{\Phi} B}{(n+1)\rho}
\ee
where $\beta_\gamma= \frac{m_{\rm Pl}}{M_\gamma}$.
Notice that this is independent of the energy of the photon beam and depends essentially on the photon flux.

Let us now consider an experiment in which chameleons are created in a laboratory vacuum before bouncing off a wall; the density is assumed to vary from
$\rho_b$ to $\rho_c$. When approaching the wall, the chameleon mass interpolates between the vacuum one in a sparse region with a density $\rho_b$ to the one in a dense medium with density $\rho_c$. In the vicinity of the wall, we have $\phi\approx \phi_c$ close enough to the wall.
As the chameleon gets closer, its wave function evolves as
\be
\delta \phi (x)= \sqrt{\frac{\pi k x}{\alpha}} J_\alpha (kx) \delta\phi_\infty
\ee
where
\be
\alpha=\vert \frac{2+3n}{4+2n}\vert
\ee
and $\delta\phi_\infty$ is the amplitude far from the wall~\cite{Brax_etal_2007b}. The wall is located at $x=0$. Similarly the background value is given by
\be
\phi_0 (x)= \phi_W (1+ \frac{m_W\vert 2+n\vert}{\sqrt{2n(n+1)}}x)^{2/(n+2)}
\ee
where $\phi_W= (1+1/n) \phi_b$ and $m_W=m_{\rm eff}(\phi_W)= \sqrt{n(n+1) \Lambda^{n+4}\phi_W^{-2-n}}$.
For small $kx$, we have
\be
\frac{\delta\phi}{\phi_0}(x)\approx \sqrt{\frac{\pi}{\alpha}} \frac{\delta\phi_\infty}{\Gamma (1+\alpha)} \alpha^{-\alpha} (\sqrt{\frac{\Lambda^{n+4}}{\alpha}}\frac{2+n}{k})^{-2/(n+2)}
(kx)^{2n/(n+2)}.
\ee
The perturbation is always smaller close to the wall than in the bulk.

The result (\ref{e:rate_p})  applies to frames in which the chameleon particles are all non-relativistic.  By summing this expression over $p$ and $q$ we may explicitly compute the fragmentation rate within a chameleon coherent state in experiments such as CHASE, assuming that the scatter $\sigma$ in momentum is smaller than the mass.  This formula should also apply during a wall reflection, since the large chameleon mass near the wall means that the particles are approximately non-relativistic in the laboratory frame around the time of the bounce.
Hence we can estimate the fragmentation rate for cavity experiments.


In practice we have
\be
\frac{\delta\phi}{\phi_0}= \frac{6\cdot 10^{-17}{\rm g\cdot cm^{-3}}}{\rho} \frac{2\beta_\gamma}{(n+1)\beta}
\ee
for a beam of power $10^6\ {\rm W}$, a cross section of $1\  {\rm cm^2}$ and a magnetic field of $1\ {\rm T}$. The density of a gas at a pressure
$10^{-8}\ {\rm mbar}$ and temperature $1$ K for instance corresponds to a density of order $10^{-13}$ g/cm$^3$ and  $\delta \phi/ \phi_0 \ll 1$.

In a cavity experiment of the finite size $R \approx 3$~cm like for the CHASE oscillation chamber, the chameleon has a lower bound on its mass $m_R \sim 10^{-5}$~eV and an upper bound
\begin{equation}
\phi_R = \Lambda (n(n+1)\Lambda^2 / m_R^{2})^{1/(n+2)}
\label{e:phi_R}
\end{equation}
on the field VEV.  The nonzero photon plasma frequency would impose a similar lower bound on the difference between chameleon and photon masses, similarly decreasing mixing.  The mixing angle is then $\theta = k \bgam B \Mpl^{-1} m_R^{-2}$ for $\bgam = \Mpl / M_\gamma$, where $\Mpl = 2.4\times 10^{18}$~GeV is the reduced Planck mass.
Using this value for the mixing angle, $P_\gamma = 3$~W for the power, and $S=1$~cm$^2$ for the area of the beam, we find $\delta\phi / \phi_R = 3 \times 10^{-17} \bgam (B/1 \textrm{ T})$ for $n=1$.  As CHASE probed the largest $\bgam$ using magnetic fields much smaller than $1$~Tesla,  $\delta\phi/\phi_R$ is many orders of magnitude smaller than unity in CHASE.  Therefore, fragmentation due to momentum dispersion $\sigma \ll m_0$ and wall collisions is negligible in afterglow experiments.

\subsection{Two-state fragmentation and implications for CHASE}

The remaining fragmentation process is the two-state interaction, similar to the two-particle interaction  discussed in~\cite{Upadhye_Steffen_Chou_2012}.    A small $\delphi/\phi_0$ can balance a large $\sigma/\phi_0$, making the overall rate convergent.  For example, assume $n=1$, $\bgam=10^{11}$, $m_0=10^{-5}$~eV, and $\sigma = 2.33$~eV as in CHASE.  Then~(\ref{e:phi_R}) implies $\phi_0 \sim 0.1$~eV, and for the $B=5$~Tesla run, we have $\delphi_0/\phi_0 \sim 10^{-5}$.  The fragmentation rate~(\ref{e:rate_large_sigma}) should be largest at small $p$ and $q$, so consider $q=5$ and $p=3$, the single-particle-production process with the lowest $p$ and $q$.  Assuming $c_q\sim 1$,~(\ref{e:rate_large_sigma}) gives a fragmentation rate $\sim 10^{-6}$~Hz for the volume of a photon pulse in CHASE, which is many orders of magnitude below the chameleon decay rate for this model.

\begin{figure}[tb]
\begin{center}
\includegraphics[angle=0,width=2.9in]{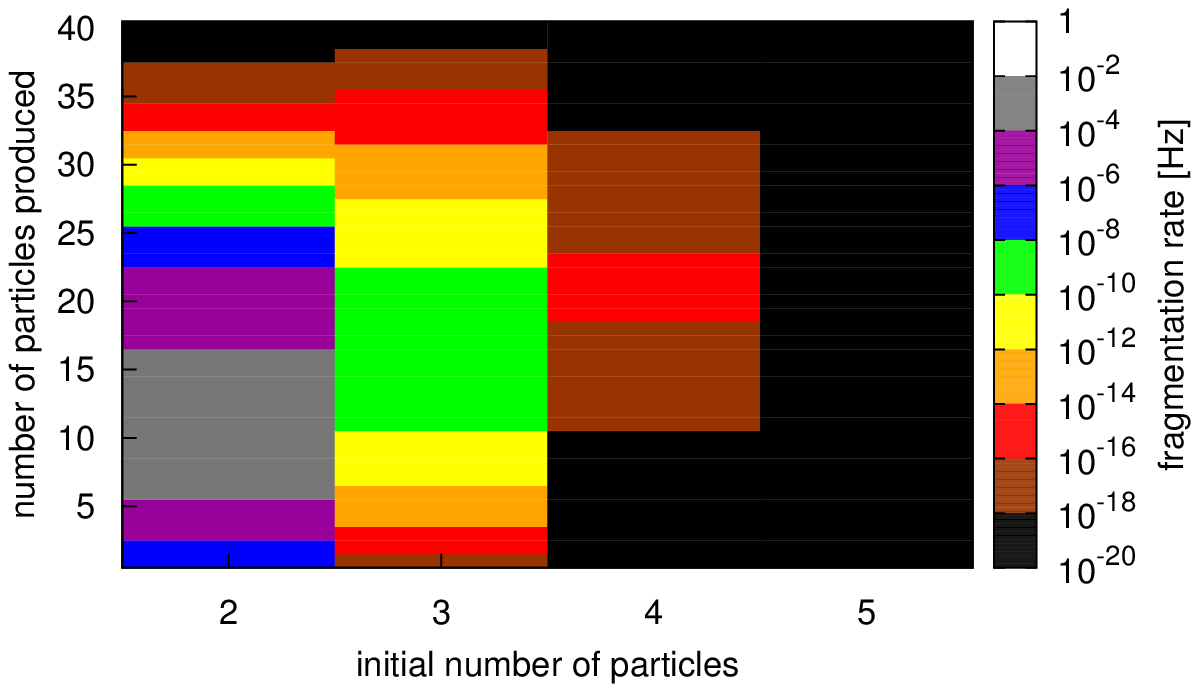}
\includegraphics[angle=0,width=2.9in]{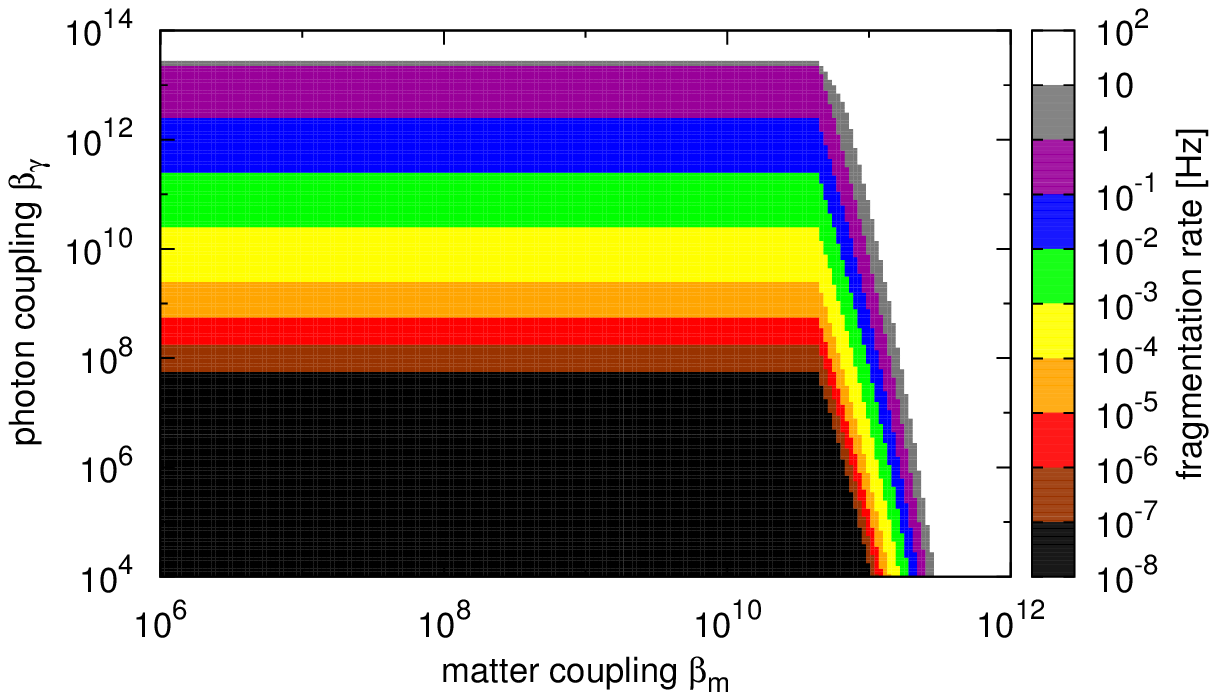}
\caption{Fragmentation rate in CHASE for an $n=1$ potential in a $5$~Tesla magnetic field. (Left)~The rate~(\ref{e:rate_large_sigma}) for $\bgam=\bmat=10^{10}$, as a function of the initial number of particles $q-p$ and the number of particles produced $2p-q$, is strongly peaked.  (Right)~The rate maximized over $p$ and $q$ becomes large for $\bmat > 10^{11}$. \label{f:chase_n1}}
\end{center}
\end{figure}

We now carry out a systematic estimate of the fragmentation rate in CHASE, and its implications for CHASE constraints.  The most commonly-considered potentials are $\lambda \phi^4/4!$, corresponding to $n=-4$, and $\Lambda^4 \exp(\Lambda/\phi)$, approximately corresponding to $n=1$.  The CHASE results~\cite{Steffen_etal_2010,Upadhye_Steffen_Chou_2012} emphasize constraints on
({\emph{i}}) a model-independent photon-coupled particle which is trapped in the CHASE chamber and does not fragment;
({\emph{ii}}) a photon-coupled chameleon with $n=1$, representative of positive-$n$ models; and
({\emph{iii}}) a photon-coupled chameleon with $n=-4$, representative of negative-$n$ models.
We will focus on $n=1$ and $n=-4$ here.  Since $n=-4$ has already been excluded by the Casimir constraints of~\cite{Brax_etal_2007c,Brax_etal_2010b}, $n=1$ is the most interesting case to consider.

As we will show, the CHASE excluded region for $n=1$ chameleons is essentially unaffected by fragmentation.  As shown in Fig.~\ref{f:chase_n1}~(Left), the rate~(\ref{e:rate_large_sigma}) as a function of $q$ and $p$ is sharply peaked, with only a few other terms within an order of magnitude of the largest term.  Thus we are justified in approximating the total fragmentation rate by the maximum of~(\ref{e:rate_large_sigma}) over $q$ and $p$.  Figure~\ref{f:chase_n1}~(Right) shows this maximum rate as a function of the chameleon matter and photon couplings.  A comparison of the low-$\bmat$ fragmentation rate with Fig.~$27$ of~\cite{Upadhye_Steffen_Chou_2012} shows that the fragmentation rate will have a negligible effect on the CHASE constraints.  The $B=5$~Tesla run is relevant for $\bgam < 10^{12}$; larger $\bgam$ are excluded by lower-$B$ runs.  Meanwhile, fragmentation does not have a significant effect on constraints until its rate exceeds $\sim 1$~Hz, which for $B=5$~Tesla is approximately $\bgam > 10^{13}$.  At still greater $\bgam$, CHASE constraints come from lower and lower magnetic field data runs, all the way down to $B=0.05$~Tesla.  This low-$B$ run, in turn, overlaps considerably with the collider constraints of~\cite{Brax_etal_2009}.  Thus the overlap of many data runs with $B$ varying over two orders of magnitude means that fragmentation has only a negligible effect on the CHASE excluded region for $n=1$ models and $\bmat \lesssim 10^{11}$.  Though constraints for $\bmat \gtrsim 10^{11}$ are somewhat weakened by fragmentation, experiments using atoms and cold neutrons~\cite{Brax_Burrage_2011,Brax_Pignol_2011,Nesvizhevsky_etal_2002} also exclude that parameter region.  More generally, for flat potentials $n \lesssim 1$, the background chameleon field is large in the CHASE vacuum chamber, $\phi_0 \gg \Lambda$, so that a small $\delta\phi_0 / \phi_0$ suppresses the fragmentation rate.  In more steeply-falling potentials $n \gtrsim 2$, $\phi_0$ is small and fragmentation can be significant.  We find that for $n \geq 2$, fragmentation reduces the expected chameleon population by several orders of magnitude, weakening constraints.

\begin{figure}[tb]
\begin{center}
\includegraphics[angle=0,width=3.14in]{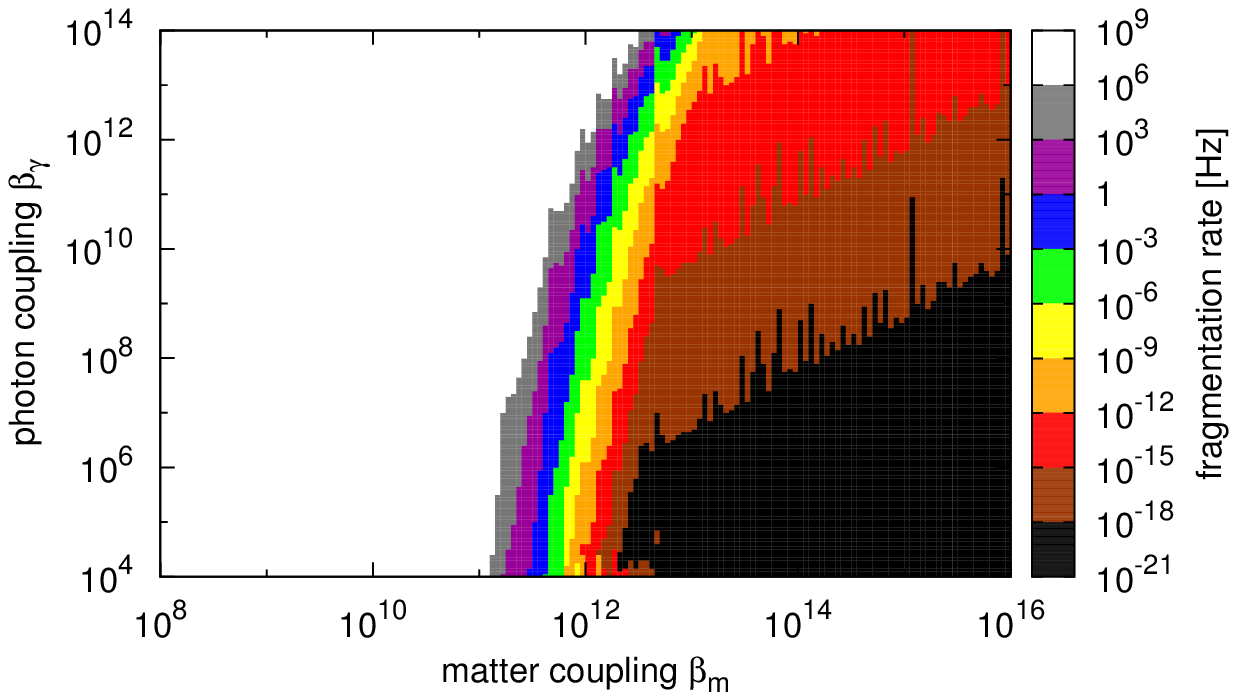}
\includegraphics[angle=0,width=2.66in]{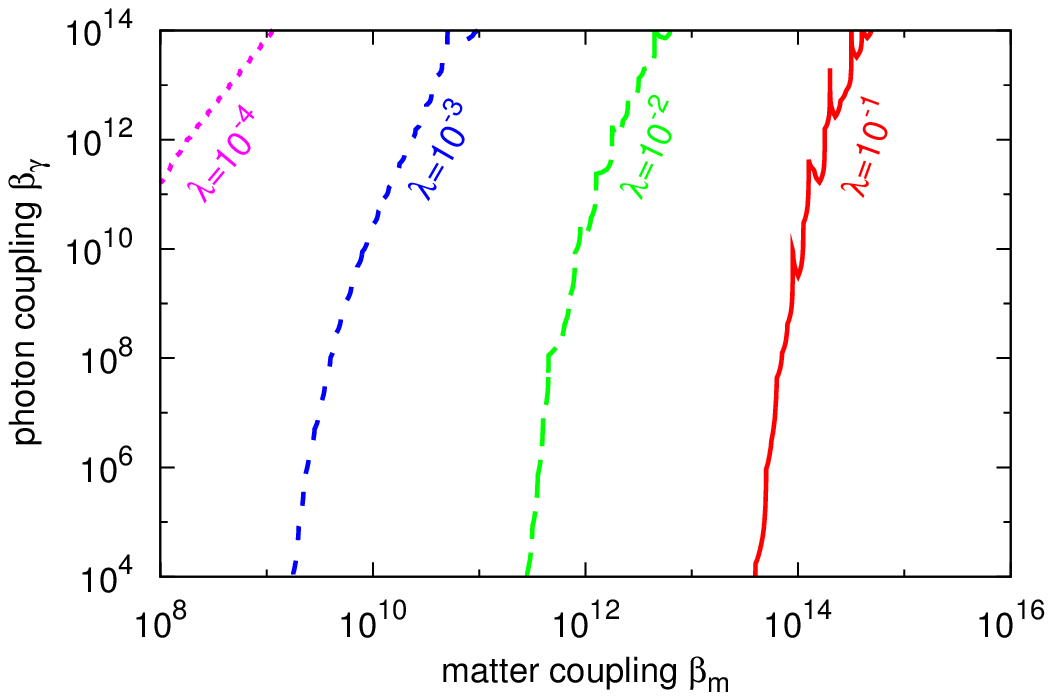}
\caption{Fragmentation rate in CHASE for $V(\phi) = \lambda\phi^4/4!$.  (Left)~Rate for $\lambda=10^{-2}$.  (Right)~For each $\lambda$, the fragmentation rate is below $1$~Hz for models below and to the right of the corresponding curve.\label{f:n-4}}
\end{center}
\end{figure}

The $n = -4$ fragmentation calculation is somewhat trickier.  If we consider only the tree-level potential, then the Taylor expansion~(\ref{e:taylor_expansion}) for $V(\phi_0 + \delta\phi)$ truncates at fourth order, while all contributions to fragmentation come from terms of order five and higher.  Our approach is to expand the one-loop effective potential instead of the tree-level potential; for small $\lambda$, this should be a reasonable estimate.  We find that the fifth derivative of the potential is $V^{(5)}(\phi) = 3\lambda^2/(16\pi^2\phi)$, hence
\begin{equation}
c_q
=
(-1)^{q+1}
\frac{3 \lambda^2}{16\pi^2}
\frac{(q-5)!}{q!}
\label{e:cq_phi4}
\end{equation}
for $q \geq 5$ in~(\ref{e:taylor_expansion}).  Figure~\ref{f:n-4} shows the corresponding fragmentation rate in CHASE.  Since CHASE excludes a range of $\bgam$ for $\bmat \gtrsim 10^{11}$ in this range of $\lambda$, it is evident from Fig.~\ref{f:n-4}~(Right) that fragmentation in CHASE is negligible for $\lambda \leq 10^{-3}$ and small for $\lambda = 10^{-2}$.  This conclusion is in qualitative agreement with the CHASE analysis~\cite{Upadhye_Steffen_Chou_2012}.  Since $n=-4$ is the only potential for which Ref.~\cite{Upadhye_Steffen_Chou_2012} computed fragmentation rates, this case is an important cross-check.

\begin{figure}[tb]
\begin{center}
\includegraphics[angle=0,width=2.9in]{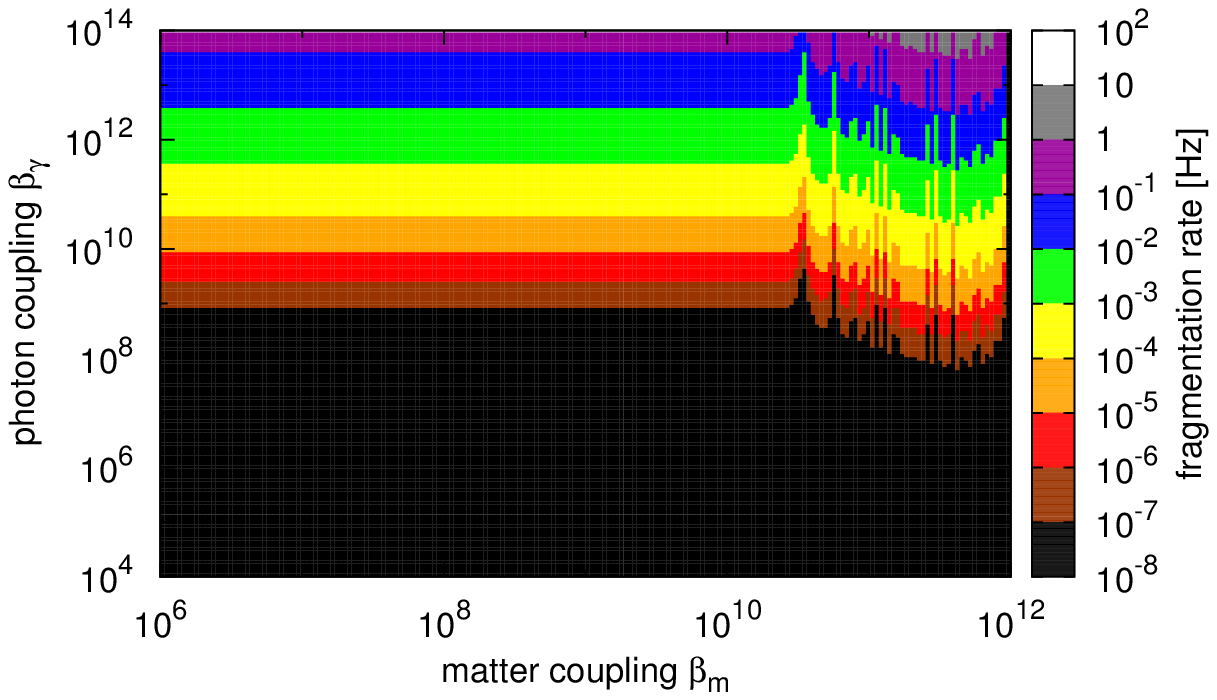}
\includegraphics[angle=0,width=2.9in]{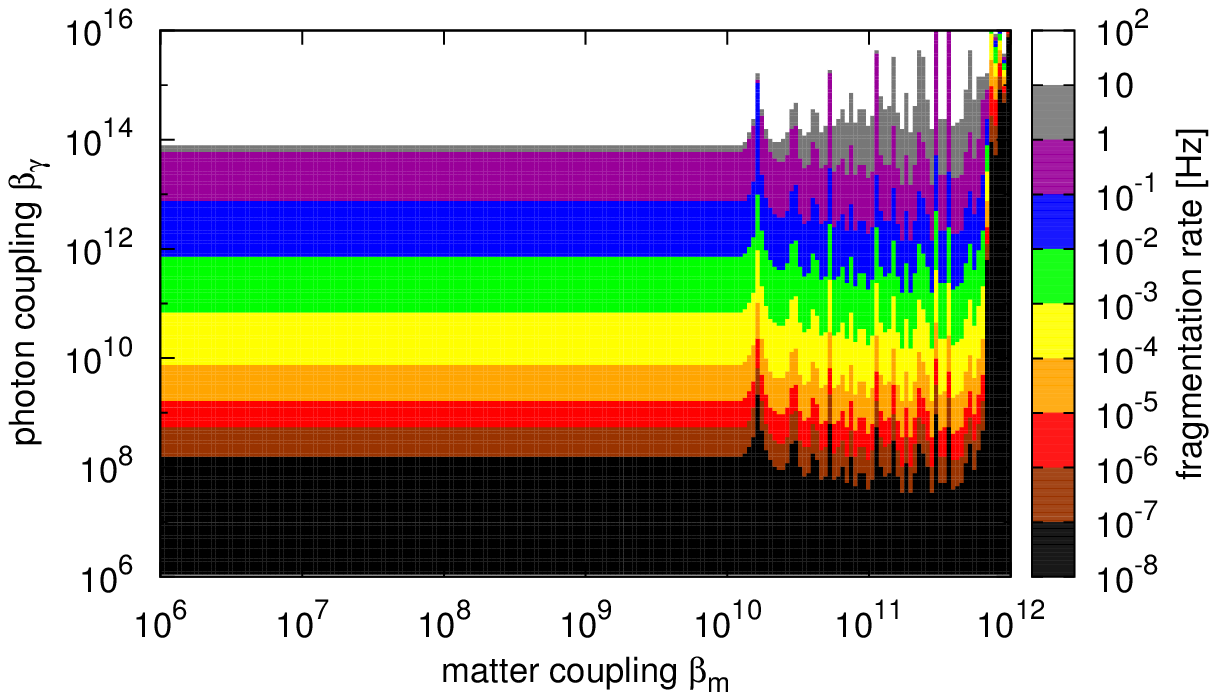}
\caption{Fragmentation rates in a hypothetical afterglow experiment with an input photon energy of $10^{-4}$~eV and a CHASE-like geometry.  (Left)~$n=2$. (Right)~$n=4$.   \label{f:k1e-4}}
\end{center}
\end{figure}

Although CHASE used a photon energy $2.33$~eV~$ \gg \Lambda$ and a photomultiplier tube (PMT) detector sensitive only to energies $\gtrsim 1$~eV, a future experiment could use an input photon energy smaller than $\Lambda$ and be sensitive to regenerated photons over a larger range of energies.  Since we do not have a specific proposal in mind, we estimate the capabilities of such an instrument by assuming a geometry identical to that of CHASE and a photon energy of $10^{-4}$~eV.  Such an experiment could push the constraints of CHASE to higher $n$.  More interestingly, for sufficiently steep potentials $n \gg 1$ the lower-energy detectors of this experiment could see the lower-energy photons produced from fragmentation products, a unique signature of scalar dark energy with a non-linear self-interaction.  Figure~\ref{f:k1e-4} shows the fragmentation rate in such an experiment for $n=2$ and $n=4$.  Since $k \ll \Lambda$, fragmentation is well-controlled even for $n=4$, though it is noticably larger than for $n=2$.  The sensitivity attainable in such an experiment is not yet known, but it is likely that for $n\geq 4$, there are models not excluded by CHASE whose fragmentation products could be detectable.

\section{Conclusion}
\label{sec:conclusion}

We have carried out a semiclassical computation in afterglow experiments of chameleon fragmentation, by which a few initial chameleon particles produce many lower-energy chameleon particles.  Our results are appropriate to coherent chameleon states, such as would be produced through oscillation from a laser pulse containing a large number of photons.  The interaction of two such coherent pulses facilitates fragmentation processes requiring high center-of-mass energies, which dominate the total fragmentation rate.

Fragmentation is a quantum mechanical process, and we find, not surprisingly, that it is suppressed in the ``classical regime'' where:
\begin{itemize}
\item[({\emph{i}})] the oscillation amplitude $\delta\phi$ about the VEV $\phi_0$ is small, $|\delta\phi / \phi_0| \ll 1$;
\item[({\emph{ii}})] the center-of-mass momenta are small, $|k / \phi_0| \ll 1$; or,
\item[({\emph{iii}})] for inverse power law potentials $V = \Lambda^4(1 + \Lambda^n/\phi^n)$, the field-dependent potential $V-\Lambda^4$ remains below the cutoff $\Lambda^4$, that is, the VEV is large, $|\phi_0 / \Lambda| \gg 1$.
\end{itemize}
All of these contribute to the suppression of fragmentation in interactions between coherent states in afterglow experiments.  Equation~(\ref{e:rate_large_sigma}) is our estimate of the fragmentation rate due to the interaction between two coherent states.

Using these results, we have considered the implications of chameleon fragmentation for completed as well as planned afterglow experiments.  Exclusion limits set by the CHASE experiment~\cite{Steffen_etal_2010} at Fermilab in $2010$ are essentially unaffected for the most commonly-considered potentials, $n=1$ and $n=-4$.  Fragmentation rates for these two potentials are shown in Figures~\ref{f:chase_n1} and \ref{f:n-4}, respectively.  In the $n=-4$ case, our calculation of the fragmentation rate is consistent with the previous result of~\cite{Upadhye_Steffen_Chou_2012}.  Meanwhile, for sufficiently large $n$ the fragmentation rate is significant; we find that CHASE constraints are substantially weakened for $n \geq 2$.

A future afterglow experiment could extend constraints to higher $n$ by using lower photon energies, suppressing fragmentation.  Equipped with a detector sensitive to even lower energies, it could potentially detect photons generated by fragmentation products, a unique signature of non-linear self-interactions, as shown in Fig.~\ref{f:k1e-4}.  Meanwhile, further theoretical work would be required to extend our results to helioscope experiments, which search for photon-coupled chameleon particles emitted by the Sun.  Although such chameleons are produced incoherently, the large VEVs predicted in positive-power-law potentials, as well as the small oscillation amplitudes associated with single-particle production, might still keep fragmentation under control at typical solar chameleon energies $\approx 600$~eV.

\bibliographystyle{JHEP}
\bibliography{chameleon1}

\providecommand{\href}[2]{#2}\begingroup\raggedright\begin{thebibliography}{10}

\bibitem{Suzuki_etal_2012}
N.~Suzuki et~al. {\em Astrophys.~J.} {\bf 746} (2012) 85.

\bibitem{Ade_etal_2013xvi}
P.~A.~R. Ade et~al. e-Print arXiv:1303.5076.

\bibitem{Anderson_etal_2013}
L.~Anderson et~al. {\em Mon.~Not.~R.~Astron.~Soc.} {\bf 427} (2013) 3435.

\bibitem{Ratra_Peebles_1988}
B.~Ratra and P.~J.~E. Peebles {\em Phys. Rev. D} {\bf 37} (1988) 3406.

\bibitem{Peebles_Ratra_1988}
P.~J.~E. Peebles and B.~Ratra {\em Ap. J. Lett.} {\bf 325} (1988) 17.

\bibitem{Reuter_Wetterich_1987}
M.~Reuter and C.~Wetterich {\em Phys. Lett. B} {\bf 188} (1987) 38.

\bibitem{Wetterich_1988}
C.~Wetterich {\em Nucl. Phys. B} {\bf 302} (1988) 668.

\bibitem{Khoury_Weltman_2004a}
J.~Khoury and A.~Weltman {\em Phys.~Rev.~Lett.} {\bf 93} (2004) 171104.

\bibitem{Khoury_Weltman_2004b}
J.~Khoury and A.~Weltman {\em Phys. Rev. D} {\bf 69} (2004) 044026.

\bibitem{Brax_etal_2004}
P.~Brax, C.~van~de Bruck, A.-C. Davis, J.~Khoury, and A.~Weltman {\em Phys.
  Rev. D} {\bf 70} (2004) 123518.

\bibitem{Brax:2010gi}
P.~Brax, C.~van~de Bruck, A.-C. Davis, and D.~Shaw, {\it {The Dilaton and
  Modified Gravity}},  {\em Phys.Rev.} {\bf D82} (2010) 063519,
  [\href{http://xxx.lanl.gov/abs/1005.3735}{{\tt arXiv:1005.3735}}].

\bibitem{Damour:1994zq}
T.~Damour and A.~M. Polyakov, {\it {The String dilaton and a least coupling
  principle}},  {\em Nucl. Phys.} {\bf B423} (1994) 532--558,
  [\href{http://xxx.lanl.gov/abs/hep-th/9401069}{{\tt hep-th/9401069}}].

\bibitem{Pietroni:2005pv}
M.~Pietroni, {\it {Dark energy condensation}},  {\em Phys.Rev.} {\bf D72}
  (2005) 043535, [\href{http://xxx.lanl.gov/abs/astro-ph/0505615}{{\tt
  astro-ph/0505615}}].

\bibitem{Olive_Pospelov_2007}
K.~A. Olive and M.~Pospelov {\em Phys.Rev.} {\bf D77} (2008) 043524.

\bibitem{Hinterbichler_Khoury_2010}
K.~Hinterbichler and J.~Khoury {\em Phys.~Rev.~Lett.} {\bf 104} (2010) 231301.

\bibitem{Nicolis_Rattazzi_Trincherini_2008}
A.~Nicolis, R.~Rattazzi, and E.~Trincherini {\em Phys.Rev.} {\bf D79} (2009)
  064036.

\bibitem{Kapner_etal_2007}
D.~J. Kapner, T.~S. Cook, E.~G. Adelberger, J.~H. Gundlach, B.~R. Heckel, C.~D.
  Hoyle, and H.~E. Swanson {\em Phys.~Rev.~Lett.} {\bf 98} (2007) 021101.
  e-Print arXiv:hep-ph/0611184.

\bibitem{Adelberger_etal_2007}
E.~G. Adelberger, B.~R. Heckel, S.~A. Hoedl, C.~D. Hoyle, D.~J. Kapner, and
  A.~Upadhye {\em Phys.~Rev.~Lett.} {\bf 98} (2007) 131104. e-Print
  arXiv:hep-ph/0611223.

\bibitem{Mota_Shaw_2006}
D.~F. Mota and D.~J. Shaw {\em Phys. Rev. Lett.} {\bf 97} (2006) 151102.

\bibitem{Mota_Shaw_2007}
D.~F. Mota and D.~J. Shaw {\em Phys. Rev. D.} {\bf 75} (2007) 063501.

\bibitem{Upadhye_Gubser_Khoury_2006}
A.~Upadhye, S.~S. Gubser, and J.~Khoury {\em Phys.~Rev.~D} {\bf 74} (2006)
  104024.

\bibitem{Brax_etal_2007c}
P.~Brax, C.~van~de Bruck, A.~C. Davis, D.~F. Mota, and D.~J. Shaw {\em
  Phys.~Rev.~D} {\bf 76} (2007) 124034. e-Print arXiv:0709.2075.

\bibitem{Brax_etal_2009}
P.~Brax, C.~Burrage, A.-C. Davis, D.~Seery, and A.~Weltman {\em JHEP} {\bf
  0909} (2009) 128. e-print arXiv:0904.3002.

\bibitem{Nesvizhevsky_etal_2002}
V.~V. Nesvizhevsky et~al. {\em Nature} {\bf 415} (2002) 297.

\bibitem{Brax_Burrage_2011}
P.~Brax and C.~Burrage {\em Phys.~Rev.~D} {\bf 83} (2011) 035020.

\bibitem{Brax_Pignol_2011}
P.~Brax and G.~Pignol {\em Phys.~Rev.~Lett.} {\bf 107} (2011) 111301.

\bibitem{Upadhye_2012}
A.~Upadhye {\em Phys.~Rev.~D} {\bf 86} (2012) 102003. e-Print: arXiv:1209.0211.

\bibitem{Brax_Pignol_Roulier_2013}
P.~Brax, G.~Pignol, and D.~Roulier. e-Print: arXiv:1306.6536.

\bibitem{Gubser_Khoury_2004}
S.~S. Gubser and J.~Khoury {\em Phys.~Rev.~D} {\bf 70} (2004) 104001.

\bibitem{Adelberger_Heckel_Nelson_2003}
E.~G. Adelberger, B.~R. Heckel, and A.~E. Nelson {\em
  Ann.~Rev.~Nucl.~Part.~Sci.} {\bf 53} (2003) 77--121.

\bibitem{Hu_Sawicki_2007}
W.~Hu and I.~Sawicki {\em Phys. Rev. D} {\bf 76} (2007) 064004.

\bibitem{Upadhye_Hu_2009}
A.~Upadhye and W.~Hu {\em Phys. Rev. D} {\bf 80} (2009) 064002.

\bibitem{Babichev_Langlois_2009}
E.~Babichev and D.~Langlois {\em Phys.~Rev.~D} {\bf 80} (2009) 121501. e-Print
  arXiv:0904.1382.

\bibitem{Upadhye_Steffen_2013}
A.~Upadhye and J.~H. Steffen. e-Print: arXiv:1306.6113 (submitted to PRL).

\bibitem{Brax_Davis_2013}
P.~Brax and A.-C. Davis. e-Print: arXiv:1301.5587.

\bibitem{Oyaizu_2008}
H.~Oyaizu {\em Phys.~Rev.~D} {\bf 78} (2008) 123523.

\bibitem{Oyaizu_Lima_Hu_2008}
H.~Oyaizu, M.~Lima, and W.~Hu {\em Phys.~Rev.~D} {\bf 78} (2008) 123524.

\bibitem{Schmidt_etal_2009}
F.~Schmidt, M.~V. Lima, H.~Oyaizu, and W.~Hu {\em Phys.~Rev.~D} {\bf 79} (2009)
  083518.

\bibitem{Bernardeau_Brax_2011}
F.~Bernardeau and P.~Brax {\em JCAP} {\bf 1106} (2011) 019.

\bibitem{Cabre_etal_2012}
A.~Cabre, V.~Vikram, G.-B. Zhao, B.~Jain, and K.~Koyama {\em JCAP} {\bf 1207}
  (2012) 034.

\bibitem{Jain_Vikram_Sakstein_2012}
B.~Jain, V.~Vikram, and J.~Sakstein. e-Print: arXiv:1204.6044.

\bibitem{Brax_Valageas_2012}
P.~Brax and P.~Valageas {\em Phys.~Rev.~D} {\bf 86} (2012) 063512.

\bibitem{Li_etal_2012a}
B.~Li, G.-B. Zhao, R.~Teyssier, and K.~Koyama {\em JCAP} {\bf 1201} (2012) 051.

\bibitem{Lee_etal_2013}
J.~Lee, G.-B. Zhao, B.~Li, and K.~Koyama {\em Ap.J.} {\bf 763} (2013) 28.

\bibitem{Vikram_etal_2013}
V.~Vikram, A.~Cabre, B.~Jain, and J.~VanderPlas. e-Print: arXiv:1303.0295.

\bibitem{Chou_etal_2009}
A.~S. Chou, W.~C. Wester, A.~Baumbaugh, H.~R. Gustafson, Y.~Irizarry-Valle,
  P.~O. Mazur, J.~H. Steffen, R.~Tomlin, A.~Upadhye, A.~Weltman, X.~Yang, and
  J.~Yoo {\em Phys. Rev. Lett} {\bf 102} (2009) 030402.

\bibitem{Steffen_etal_2010}
J.~H. Steffen et~al. {\em Phys.~Rev.~Lett.} {\bf 105} (2010) 261803. ePrint:
  arXiv:1010.0988.

\bibitem{Brax_etal_2007b}
P.~Brax, C.~van~de Bruck, A.~C. Davis, D.~F. Mota, and D.~J. Shaw {\em
  Phys.~Rev.~D} {\bf 76} (2007) 085010. e-Print arXiv:0707.2801.

\bibitem{Ahlers_etal_2008}
M.~Ahlers et~al. {\em Phys. Rev. D.} {\bf 77} (2008) 015018.

\bibitem{Gies_Mota_Shaw_2008}
H.~Gies, D.~F. Mota, and D.~J. Shaw {\em Phys. Rev. D} {\bf 77} (2008) 025016.

\bibitem{Upadhye_Steffen_Weltman_2010}
A.~Upadhye, J.~H. Steffen, and A.~Weltman {\em Phys.~Rev.~D} {\bf 81} (2010)
  015013.

\bibitem{Upadhye_Steffen_Chou_2012}
A.~Upadhye, J.~H. Steffen, and A.~S. Chou {\em Phys. Rev. D} {\bf 86} (2012)
  035006.

\bibitem{Burrage_Davis_Shaw_2009}
C.~Burrage, A.-C. Davis, and D.~J. Shaw {\em Phys. Rev. D} {\bf 79} (2009)
  044028.

\bibitem{Brax_Zioutas_2010}
P.~Brax and K.~Zioutas {\em Phys.~Rev.~D} {\bf 82} (2010) 043007.

\bibitem{Brax_Lindner_Zioutas_2012}
P.~Brax, A.~Lindner, and K.~Zioutas {\em Phys.~Rev.~D} {\bf 85} (2012) 043014.

\bibitem{Fujii_1997}
Y.~Fujii {\em Mod.~Phys.~Lett.~A} {\bf 12} (1997) 371--380. e-Print:
  gr-qc/9610006.

\bibitem{Brax_Martin_2007}
P.~Brax and J.~Martin {\em Phys.~Lett.~B} {\bf 647} (2007) 320--329.

\bibitem{Hui_Nicolis_2010}
L.~Hui and A.~Nicolis {\em Phys.~Rev.~Lett.} {\bf 105} (2010) 231101.

\bibitem{Hinterbichler_Khoury_Nastase_2011}
K.~Hinterbichler, J.~Khoury, and H.~Nastase {\em JHEP} {\bf 1103} (2011),
  no.~61.

\bibitem{Upadhye_Hu_Khoury_2012}
A.~Upadhye, W.~Hu, and J.~Khoury {\em Phys. Rev. Lett} {\bf 109} (2012) 041301.

\bibitem{Erickcek_etal_2013}
A.~Erickcek, N.~Barnaby, C.~Burrage, and Z.~Huang. e-Print: arXiv:1304.0009.

\bibitem{Erickcek:2013dea}
A.~L. Erickcek, N.~Barnaby, C.~Burrage, and Z.~Huang, {\it {Chameleons in the
  Early Universe: Kicks, Rebounds, and Particle Production}},
  \href{http://xxx.lanl.gov/abs/1310.5149}{{\tt arXiv:1310.5149}}.

\bibitem{Brax_etal_2011}
P.~Brax, C.~Burrage, A.-C. Davis, D.~Seery, and A.~Weltman {\em Phys. Lett. B}
  {\bf 699} (2011) 5.

\bibitem{Weinberg_1995}
S.~Weinberg, {\em {The Quantum theory of fields. Vol. 1: Foundations}}.
\newblock 1995.

\bibitem{Brax_etal_2010b}
P.~Brax, C.~van~de Bruck, A.-C. Davis, D.~J. Shaw, and D.~Iannuzzi {\em
  Phys.~Rev.~Lett.} {\bf 104} (2010) 241101.

\end{thebibliography}\endgroup

\end{document}